\documentclass[twocolumn,showpacs,reprint,amsmath,amssymb,superscriptaddress]{revtex4}
\usepackage{amsmath} % need for subequations
\usepackage{graphicx} % for figures
\usepackage{dcolumn,bm,amsmath}
\newcommand{\TwoByTwo}[4]{\mbox{$\left(\begin{array}{cc}#1 & #2 \\ #3 & #4 \end{array}\right)$}}

\DeclareMathAlphabet\EuScript{U}{eus}{m}{n}
\SetMathAlphabet\EuScript{bold}{U}{eus}{b}{n}
\newcommand{\Bvec}{\boldsymbol{\EuScript{B}}}%
\newcommand{\Bsca}{\EuScript{B}}
\newcommand{\Evec}{\boldsymbol{\EuScript{E}}}%
\newcommand{\Esca}{\EuScript{E}}
\newcommand{\Dsca}{\EuScript{D}}
\DeclareMathOperator{\sech}{sech}

\begin{document}

\title{Zeeman-tuned rotational level-crossing spectroscopy in a diatomic free radical}

\author{S. B. Cahn}
\email[e-mail: ]{sidney.cahn@yale.edu}
\author{J. Ammon}
\author{E. Kirilov}
\altaffiliation[Present Address: ] {Universit\"{a}t Innsbruck, Institut f\"{u}r Experimentalphysik, Technikerstrasse 25/4, A-6020 Innsbruck, Austria}
\author{Y. V. Gurevich}
\author{D. Murphree}
\altaffiliation[Present Address: ] {3002 Avalon Cove Ct.\ NW, Rochester, MN 55901}
\affiliation{Department of Physics, Yale University, P.O. Box 208120,  New Haven, CT 06520}
\author{R. Paolino}
\affiliation{Department of Science, U.S. Coast Guard Academy, 31 Mohegan Ave.,  New London, CT 06320}
\author{D. A. Rahmlow}
\altaffiliation[Present Address: ] {Wyatt Technology Corporation, 6300 Hollister Ave., Santa Barbara, CA 93117}
\affiliation{Department of Physics, Yale University, P.O. Box 208120,  New Haven, CT 06520}
%\email{sidney.cahn@yale.edu,david.demille@yale.edu} %optional
\author{M. G. Kozlov}
\affiliation{Petersburg Nuclear Physics Institute, Gatchina, 188300, Russia}

\author{D. DeMille}
\email[e-mail: ]{david.demille@yale.edu}
\affiliation{Department of Physics, Yale University, P.O. Box 208120, New Haven, CT 06520}

\begin{abstract}
Rotational levels of molecular free radicals can be tuned to degeneracy using laboratory-scale magnetic fields. Because of their intrinsically narrow width, these level crossings of opposite-parity states have been proposed for use in the study of parity-violating interactions and other applications. We experimentally study a typical manifestation of this system using $^{138}$BaF. Using a Stark-mixing method for detection, we demonstrate level-crossing signals with spectral width as small as $6$ kHz. We use our data to verify the predicted lineshapes, transition dipole moments, and Stark shifts, and to precisely determine molecular magnetic g-factors. Our results constitute an initial proof-of-concept for use of this system to study nuclear spin-dependent parity violating effects.
\end{abstract}

\pacs{32.80.Ys, 12.15.Mm, 21.10.Ky}

\maketitle

%INTRO
It has been suggested that diatomic molecules could be used as a system to measure classes of parity-violating (PV) electroweak interactions that are difficult to access through other means \cite{Flambaum85,Kozlov91,DeMille08b}. The level structure of diatomic free radicals systematically makes it possible to tune states of opposite parity to near degeneracy, using a magnetic field such that the Zeeman shift of the electron spin matches the rotational splitting. Near such a level crossing, the mixing of these long-lived states due to nuclear spin-dependent (NSD) PV interactions is greatly enhanced \cite{Nguyen97}. This should make it feasible to measure small, poorly understood effects such as those due to nuclear anapole moments and axial hadronic-vector electronic electroweak couplings \cite{DeMille08b,Haxton01,Flambaum97}. This type of level crossing has also  been identified as an attractive system for quantum simulations of conical intersections \cite{Wallis09} or magnetic excitons \cite{Perez-Rios10}, and for sensitive detection of electric fields \cite{Alyabyshev12}. 

Here we report an experimental study of Zeeman-tuned rotational level crossings in $^{138}$BaF. Using an electric field pulse to induce transitions between the near-degenerate levels, we demonstrate the ability to understand and control the system with energy resolution at the kHz scale, as desired for the measurement of nuclear spin-dependent PV effects in similar systems. By measuring the magnetic field at several crossings, we extract precise values for poorly known magnetic $g$-factors; also, by studying transfer efficiency vs.\ electric field, we deduce values for electric dipole matrix elements between the crossing levels, and for off-resonant Stark shifts not previously considered in this system. 

%BACKGROUND INFO
The ground electronic state $X^2\Sigma$ of $^{138}$BaF has one unpaired electron \cite{Herzberg50}. The $^{19}$F nucleus has spin $I = 1/2$, while $^{138}$Ba is spinless. In the absence of external fields, the lowest energy levels are described by the Hamiltonian 
\begin{equation}
H_0 = B\mathbf{N}^2 + D\mathbf{N}^4 
+ \gamma\mathbf{N}\cdot\mathbf{S} + b\mathbf{I}\cdot\mathbf{S} + c(\mathbf{I}\cdot\mathbf{n})(\mathbf{S}\cdot\mathbf{n}),
\end{equation}
%\begin{eqnarray}
%H_0 = B\mathbf{N}^2 + D_c\mathbf{N}^4  \nonumber \\
%+ \gamma\mathbf{N}\cdot\mathbf{S} + b\mathbf{I}\cdot\mathbf{S} + c(\mathbf{I}\cdot\mathbf{n})(\mathbf{S}\cdot\mathbf{n}),
%\end{eqnarray}
where $N$ is the rotational angular momentum, $S=1/2$ is the electron spin, and $\mathbf{n}$ is a unit vector along the internuclear axis ($\hbar = 1$ throughout) \cite{Childs92,Brown03}. All parameters of $H_0$ have been precisely measured \cite{Ryzlewicz80,Ryzlewicz82,Ernst86}. The rotational constant $B$ is much larger than the spin-rotation (SR) constant $\gamma$, the hyperfine (HF) constants $b$ and $c$, and the centrifugal correction constant $D$; thus $N$ is a good quantum number, with eigenstates of energy $E_N   \approx  B N(N  +  1)$ and parity $P   =   (-1)^N$.

We use a magnetic field $\Bvec = \Bsca\hat{z}$ to Zeeman-shift sublevels of the $N^P = 0^+$ and $1^-$ manifolds of states to near degeneracy. We write the effective Zeeman Hamiltonian as \cite{Weltner83}
\begin{eqnarray}
H_Z = -g_{\perp} \mu_B \mathbf{S} \cdot \Bvec -(g_{\parallel}-g_{\perp})\mu_B (\mathbf{S} \cdot\mathbf{n})(\Bvec \cdot\mathbf{n}) \nonumber \\
           -g_I \mu_N \mathbf{I} \cdot \Bvec -g_{rot}\mu_N \mathbf{N} \cdot \Bvec, 
\end{eqnarray}
where  $\mu_B (\mu_N)$ is the Bohr (nuclear) magneton; $g_{\parallel}  \cong  -2.002$; $(g_{\parallel}  -  g_{\perp}) \cong -\gamma/(2B) = -0.00625$ (Curl equation) \cite{Curl65,Weltner83}; $g_I = 5.258$ for $^{19}$F \cite{Stone05}; and $g_{rot}$ is previously unknown for BaF.  Since $B  \gg  \gamma, b, c$, the $\Bsca$-field necessary to bridge the rotational energy $E_1  -  E_0  \approx  2 B$ is large enough to strongly decouple $\mathbf{S}$ from $\mathbf{I}$ and $\mathbf{N}$. We hence write the molecular states in the decoupled basis $|N,m_N\rangle |S,m_S\rangle |I,m_I\rangle$, which are good approximations to the energy eigenstates near the crossings. Zeeman shifts are dominated by the $g_\perp$ term, so that opposite-parity levels $|\psi^+_{\uparrow}(m_N  =  0, m_I)\rangle  \cong  |0,0\rangle |\frac{1}{2},\frac{1}{2}\rangle |\frac{1}{2},m_I\rangle$ and $|\psi^-_{\downarrow}(m_N', m_I')\rangle  \cong  |1,m_N'\rangle |\frac{1}{2},-\frac{1}{2}\rangle |\frac{1}{2},m_I'\rangle$ are degenerate under $H_0+H_Z$ when $\Bsca =\Bsca_0 \approx B/\mu_B \sim 0.5$ T in $^{138}$BaF. Level crossings between pairs of mixing states with different values of $(m_N,m_I)$ and $(m^\prime_N,m^\prime _I)$ occur at different values of $\Bsca_0$, because of energy differences in the sublevels due to HF and SR terms in $H_0$.   Level crossings in $^{138}$BaF are depicted in Fig.\ 1.

\begin{figure}
\centering
\includegraphics[width=83mm]{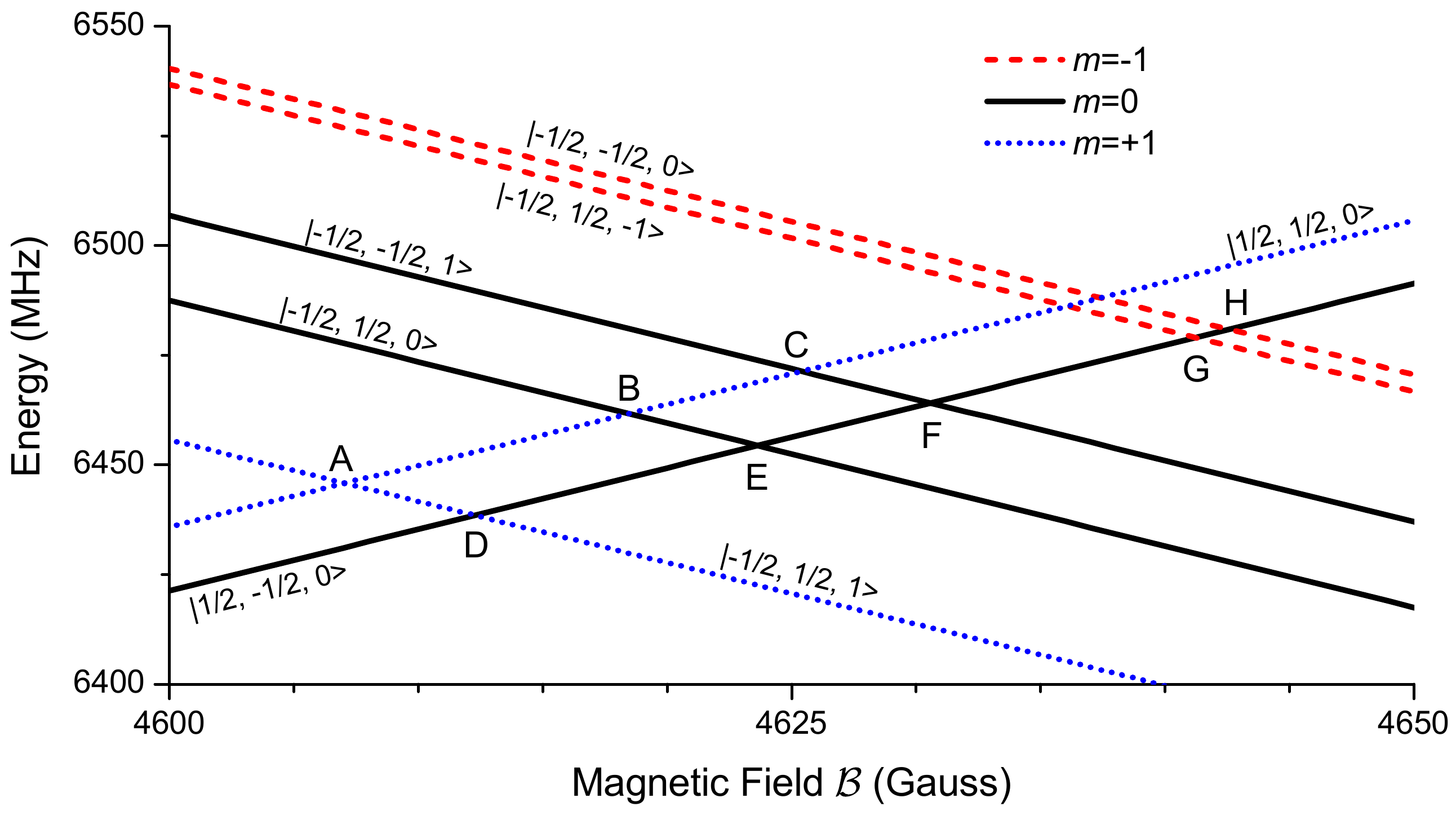}
\caption{(color online) Level crossings in $^{138}$BaF. 
Up-(down-) sloping levels belong to the even-(odd-) parity
$N=0$ $(N=1)$ rotational level. Kets label the approximate
quantum numbers $|m_S, m_I, m_N\rangle$.  Letters label each
crossing where levels can be mixed via the Stark effect.}\label{Fig1}
\end{figure}

We use an electric field $\Evec$ to mix the nearly-degenerate opposite-parity levels. The effective Stark Hamiltonian is $H_S = -\Dsca \Evec  \cdot  \mathbf{n}$, where $\Dsca = 3.170(3)$ D is the $X^2\Sigma$ state dipole moment \cite{Ernst86}. This term couples basis states with $N^\prime = N \pm 1$ and $m_S^\prime = m_S,~ m_I^\prime = m_I$. The matrix element $d\Esca_j \equiv \langle\psi^-_{\downarrow} |H_S| \psi^+_{\uparrow} \rangle $ describes the coupling between nearly-degenerate states; here the cylindrical component $j   =  z ~(j  =  \rho,\phi)$ is relevant when 
$\Delta m = m^\prime  -    m = 0 (\pm 1)$,
%$ m^\prime   =   m ~(m^\prime  =  m \pm 1)$ , 
where $m \equiv m_N + m_S + m_I$. The values of $d$ are nominally zero, since $[\mathbf{n},\mathbf{S}]  =  0$ and $m_S' \neq m_S$ for the basis states that approximately describe the degenerate levels. However, the HF and SR terms in $H_0$ and the $g_{\parallel} - g_{\perp}$ term in $H_Z$ cause a small mixture of basis states with different values of $m_S$ into the crossing levels. The induced values of $d$ can be estimated perturbatively, or calculated by diagonalizing $H_0+H_Z$ over a sufficiently large subspace (we include $N   \leq  6$). Typically $d \sim \eta \Dsca$, where $\eta \sim (\gamma,b,c)/B \ll 1$.

Near a level crossing, we describe the near-degenerate levels with the wavefunction
$ |\psi(t)\rangle = c_+(t)|\psi^+_{\uparrow}\rangle + e^{-i\Delta t}c_-(t)|\psi^-_{\downarrow}\rangle$
and the effective Hamiltonian $H_{\pm}$:
\begin{equation}
H_{\pm} = \TwoByTwo{\Delta_0(\Bsca)}{d\Esca_j}{d\Esca_j}{-\frac{\alpha}{2} \Esca^2}.
\end{equation}
Here $\Delta_0(\Bsca) \propto \Bsca-\Bsca_0$ is the small detuning from exact degeneracy under $H+H_Z$, and $\alpha$ is a small differential polarizability of the near-degenerate pair. The latter term, not considered in previous work on this system, arises from Stark-induced mixing with distant levels (outside this subspace). It is closely analogous in form and effect to the AC Stark shift that perturbs spectral lines of weak transitions used e.g. in atomic PV experiments \cite{Wieman87,Stalnaker06}. In this paper, we ignore small off-diagonal matrix elements $iW$ due to PV interactions in $H_{\pm}$ \cite{DeMille08b}.

Consider a system with initial state $c_+(0)  =  0$. An electric field pulse $\Esca_j(t)~(0  <  t  <  T)$ mixes the levels, leading to a nonzero population $P_+$ of the originally empty state: $P_+  =  |c_+(T)|^2$. This population constitutes the signal in our experiments. In certain cases, analytic solutions for $c_+(T)$ can be derived \cite{Hioe85}. One example, important for our studies, is when $\alpha \rightarrow 0$ and $\Delta = \mathrm{const.}$, and in addition $P_+   \ll  1$ so that 1$^{\mathrm{st}}$-order time-dependent perturbation theory applies.  In this case, $c_+(T; \Delta) \approx \int_0^T e^{-i\Delta t’} d\Esca_j(t’) dt’$. Hence the spectral lineshape $P_+(\Delta)$, found by tuning $\Bsca$ around $\Bsca_0$, is determined simply by the Fourier transform of the $\Esca$-field pulse. Another simple case arises for $\alpha  =  0$ at exact resonance, $\Delta  =  0$. Here, the population transfer is given by the Rabi flopping formula: $P_+ = \sin^2(\Theta/2)$, where $\Theta = \int_0^T 2d\Esca_j(t’) dt’$ is the effective pulse area \cite{Allen87}. We use this behavior to measure the values of $d$.
%
%
% Analytic solutions available in some cases
%First: Delta = const.
%Small transfer, no Stark shift: line shape = Fourier transform of E-field pulse. Example = sinusoidal E-field discussed in earlier paper for PV measurement
%Here: Gaussian or sech; deriv of Gaussian or sech*tanh; what about case with multiple rings?
%Complications from non-constant Delta from e.g. off-resonant Stark [also B-field curvature…?]: Airy
%Large E-field, on resonance: Rabi flopping.
%

%EXPERIMENTAL SYSTEM

%Near the source of a pulsed molecular beam of BaF, molecules cross a cw laser beam tuned to excite one $|\psi^+_{\uparrow}(m_N  =  0, m_I)\rangle$ sublevel to the short-lived $A ^2\Pi_{1/2}$ state \cite{Effantin90}; optical pumping depletes the targeted sublevel. 
%After passing along the axis $\hat{z}$ of a superconducting (SC) solenoid magnet, the molecules are excited on the same transition by another laser beam.  
%The molecules then traverse the main interaction region (IR), which is inside a superconducting (SC) solenoid magnet and collinear with its axis $\hat{z}$.  After exiting the magnet, another laser beam excites the molecules on the same transition.
%Here, $\approx 260$ cm from the source, an additional laser beam, tuned to the $A  -  D ^2\Sigma^+$ transition \cite{Effantin90}, is overlapped with the first beam.
Before entering the main interaction region (IR) where $\Bsca$- and $\Esca$-fields are applied, BaF molecules in a pulsed molecular beam cross a cw laser beam.  This laser serves to deplete one  $|\psi^+_{\uparrow}(m_N  =  0, m_I)\rangle$ sublevel, by optical pumping via the short-lived $A ^2\Pi_{1/2}$ state \cite{Effantin90} state.  The same transition is excited in a detection region downstream from the IR, $\approx 260$ cm from the molecular beam source, by a second laser beam.  This excitation beam is also overlapped with a third laser beam, tuned to the  $A  -  D ^2\Sigma^+$ transition \cite{Effantin90}.  Molecules entering in the $|\psi^+_{\uparrow}(m_N  =  0, m_I)\rangle$ state are excited to the $D$ state and subsequently decay to the $X$ state, emitting fluorescence that is counted with a photomultiplier. The molecular beam is collimated to radius $\rho_{max}  \approx  0.63$ cm. The depletion laser beam can be shuttered; under this condition the signal is $N_0 \approx $30-100 counts/pulse, at $\mathcal{R} = 5$-10 Hz rep.\ rate. The molecules have mean velocity $\bar{v} = 616$ m/s, with FWHM spread $\delta v/\bar{v} = 7\%$. 

\begin{figure*}
\centering
\includegraphics[width=180mm]{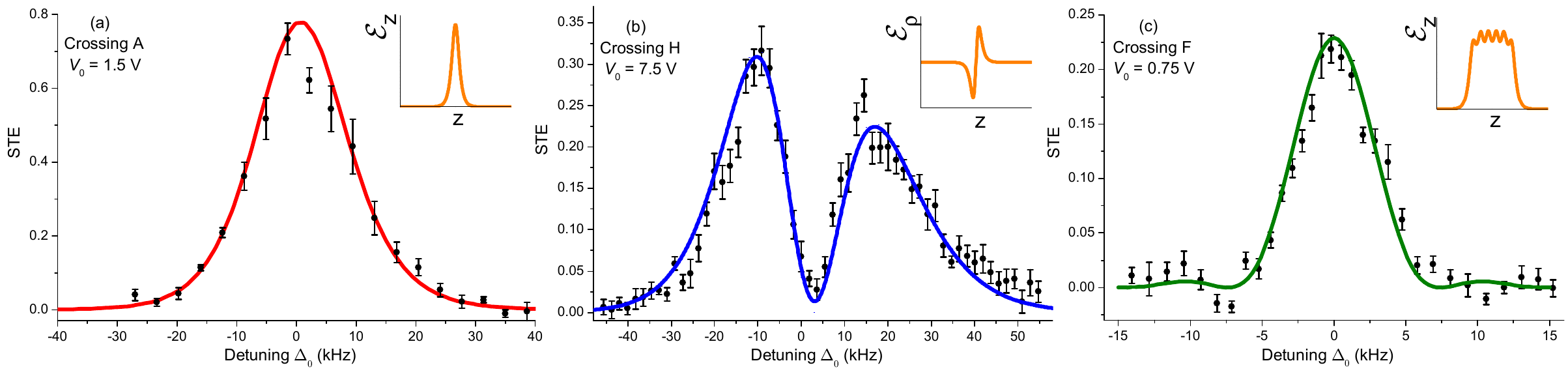}
\caption{(color online) Level crossing lineshape data (points) and predictions (curves).  
a) Single step potential and $\Esca_z$-field at a $\Delta m = 0$ transition. b) Single step potential and $\Esca_\rho$-field at a $\Delta m = 1$ transition.   c) 7-step linear potential and $\Esca_z$-field at a $\Delta m = 0$ transition.  Insets show the relevant component of $\Evec$ vs.\ $z$ position, with identical $z$ scales.  Predicted lineshapes include effects of molecule distributions in $\rho$ and $v$, and for (b) only, IR centration and the polarizability $\alpha$ (see Fig.\ 4).}\label{Fig2}
\end{figure*}

A superconducting (SC) solenoid provides the $\Bsca$-field in the IR.  The field is shimmed for homogeneity using a set of 5 SC and 14 room temperature (RT) gradient coils. Initially, the field is mapped with an array of 32 NMR probes \cite{Murphree07} surrounding the IR; each measures $\Bsca$ with precision $\delta\Bsca/\Bsca \sim 10^{-8}$. Next, currents are applied to the shim coils to minimize the r.m.s.\ deviation among the probes. Final adjustments are made using data from a similar probe translated along the molecular beam path (while not under vacuum). Afterwards, typically $\delta\Bsca/\Bsca \lesssim 10^{-7}$ over the IR volume, so that $\delta\Delta_0 \lesssim 1$ kHz. One RT coil provides a uniform field used to tune $\Bsca$; its value is monitored continuously with one probe from the array.

The $\Esca$-field in the IR is generated by voltages applied to $R=1.26(1)$ cm radius tubular electrodes centered on and stacked along the magnet axis $\hat{z}$. The electrodes at the ends of the IR are long unbroken tubes, so that $\Esca = 0$ except in a region of length $L_\Esca \lesssim 8$ cm around the center of the magnet.  The field distribution $\Evec(\rho,z)$ can be found analytically from the applied voltages $V(R,z)$. 

With the depletion laser present, the probe signal nominally appears only due to Stark-induced mixing with the thermally-populated $|\psi^-_{\downarrow}(m_N^\prime  =  0, m_I^\prime)\rangle$ state. In practice, there is a small residual signal (5-10\% of $N_0$) even with $\Esca  =  0$. This background is accounted for by defining $S(\Esca)$, the state transfer efficiency (STE), as 
$S(\Esca) = [N(\Esca,o)-N(0,o)]/[N(0,c)-N(0,o)]$. Here $N(\Esca,s)$ is the number of detected counts with electric field $\Esca$ and the depletion laser shutter $s$ open $(o)$ or closed $(c)$ [$N_0 = N(0,c)$].  If the molecular beam flux is constant for the various measurements, $ S(\Esca) = P_+$. To minimize the effect of drifts in the flux, we take data with $\Esca$ on/off and $s=o/c$, for groups of $10-20$ consecutive pulses.

For most data, $\Esca$ was produced by applying a single step in voltage $V$ at the center of the IR ($z  =  0$), so that $V(R, z) = V_0 \mathrm{sgn}(z)$. For this configuration, the axial $\Esca_z$-field near $\rho  =  0$ is well approximated by simple functions such as a Gaussian or a sech.  We describe the field as $\Esca_z(\rho,z) \approx \Esca_0 \sech(z/\sigma_z)$, where $\sigma_z \cong 0.479 R$ and $\Esca_0 \cong 1.348 V_0/R$. Molecules with velocity $v$ experience a temporal field pulse $\Esca(t) \approx \sech(t/\sigma_t)$, where $\sigma_t = \sigma_z/v$. Since the velocity distribution is narrow, $\sigma_t \approx \sigma_z/\bar{v}$. 
%Since the Fourier transform of a sech is also a sech, 
This results in the spectral lineshape $P_+(\Delta) = \sin^2(\Theta) \sech^2(\Delta/\sigma_\Delta)$, where $\sigma_\Delta = 2/(\pi\sigma_t)$ \cite{Rosen32}.  A typical level-crossing spectrum under these conditions is shown in Fig.\ 2a.

For the same configuration of voltages, transverse electric fields $\Esca_\rho$ are also present off-axis; for $\rho  \ll  R$, $\Esca_\rho(\rho,z) \approx (\Esca_0/2) (\rho/\sigma_z) \sech(z/\sigma_z) \tanh(z/\sigma_z)$.  Molecule trajectories in the collimated beam are very parallel to $\hat{z}$, so for each molecule $\rho \cong$ const.\ and $\Esca_\rho(t)$ has the form of a bipolar pulse, $\Esca_\rho(t) \propto \rho \sech(t/\sigma_t) \tanh(t/\sigma_t)$, whose magnitude varies widely across the ensemble of molecules.  For weak excitation, this yields a spectral lineshape $P_+(\Delta) \propto \Delta^2\sech^2(\Delta/\sigma_\Delta)$.  We observe lineshapes of this type at level crossings with $\Delta m  =  \pm 1$. With reasonable inferences about the centration of the IR relative to the collimating apertures (and hence the distribution of molecules in $\rho$), we reproduced the observed lineshape at these crossings (Fig.\ 2b). A bipolar pulse of similar form is required to measure PV interactions in this system \cite{DeMille08b}.

For measuring PV, the narrowest spectral lines are desired \cite{DeMille08b}.  To this end, we extend the $\Esca$-field region by inserting 5 electrode rings of length $L_i = 1.40$ cm between the long tube electrodes, and apply voltages in equal steps from $+V_0$ to $-V_0$ across this stack.  The resulting profile of $\Esca_z(\rho=0,z)$ is roughly a square pulse, with amplitude $\Esca_0 \approx 2V_0/(5L_i)$ and edges decaying over a distance $\delta z \approx \sigma_z$.  Fig.\ 2c shows the spectrum measured in this configuration.  The good agreement of the data with the predicted lineshape indicates that any additional broadening due to $\Bsca$-field inhomogeneities is negligible.  The FWHM linewidth $\delta \Delta \simeq 2\pi\times 6$ kHz sets the natural scale of energy resolution for this system; it corresponds to interaction time $T = 2\pi /\delta\Delta \approx 150~\mu$s $ \approx (5L_i+2\sigma_z)/\bar{v}$ for molecules in the $\Esca$-field.

\begin{figure}[b]
\centering
\includegraphics[width=83mm]{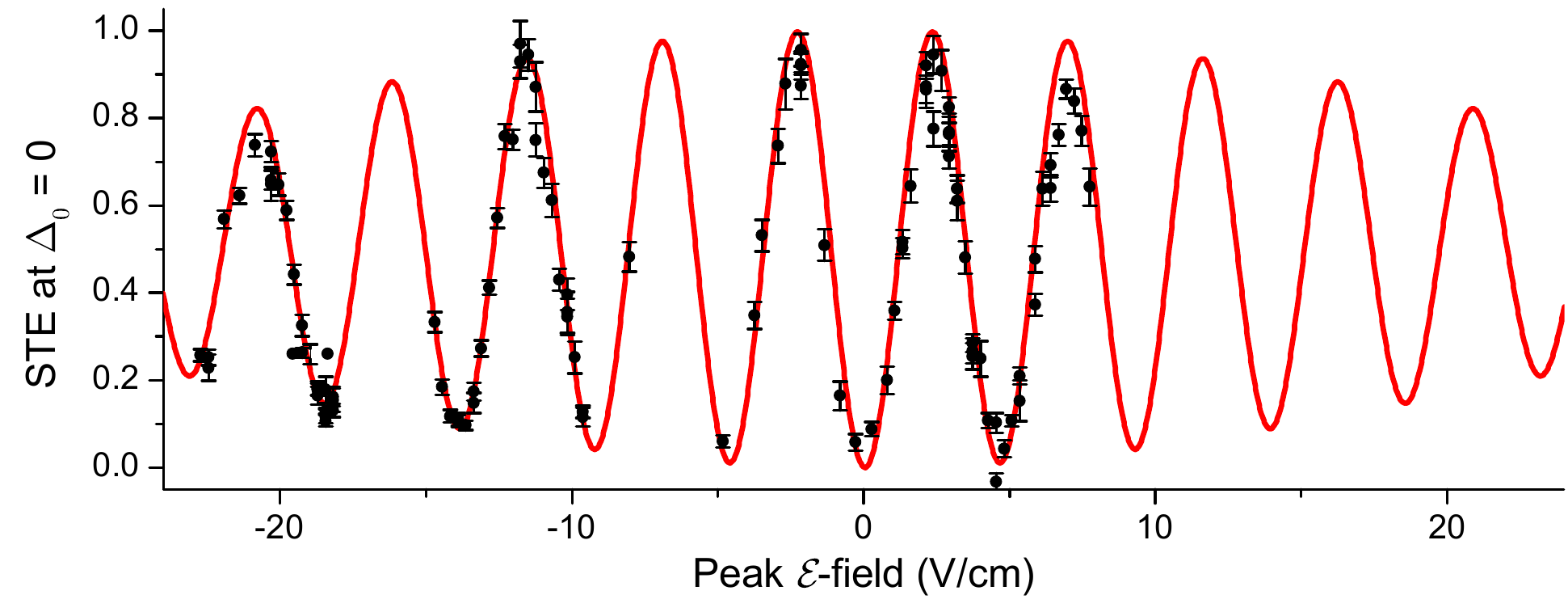}
\caption{(color online) Rabi flopping behavior on crossing A.
The fit includes the effect of the measured velocity distribution, which accounts for the contrast decay as $\mathcal{E}$ increases.  The only free parameter in the fit is the dipole matrix element, determined here as $d_A = 3.36(4)$ kHz/(V/cm).}\label{Fig3}
\end{figure}

At each crossing, we determine the transition dipole moment $d$ by mapping signal size on resonance as $\Esca$ is varied (Fig.\  3).  For $\Delta m   =  \pm 1$ crossings, the large inhomogeneity in $\Esca_\rho$-field amplitude across the molecular ensemble lead to large uncertainties in $d$.  Our measured values agree well with calculations based solely on the precisely known molecular constants (Table I).  Agreement is best on transitions with $\Delta m   =   0$ and large $|d|$, where effects due to the polarizability $\alpha$ are smallest.

At high values of $\Esca$, we frequently observed distortions of the lineshape.  These effects arise from the differential polarizability $\alpha$ in combination with the time-varying $\Esca$-field.  Values of $\alpha$ are calculated by diagonalizing $H_0+H_Z+H_S$ and extracting levels shifts vs.\ $\Esca$.  To determine $\alpha$ experimentally, we fit line shapes and positions to a numerical integration of the Schr\"{o}dinger Eqn. An analytic solution exists for a weak Gaussian $\Esca$-field pulse, 
% near resonance for the states at maximum Stark shift 
for detunings such that the maximum Stark shift brings the states near to crossing
(so only the quadratic part near the peak of the pulse contributes significantly to the state transfer).
\begin{figure}[t]
\centering
\includegraphics[width=83mm]{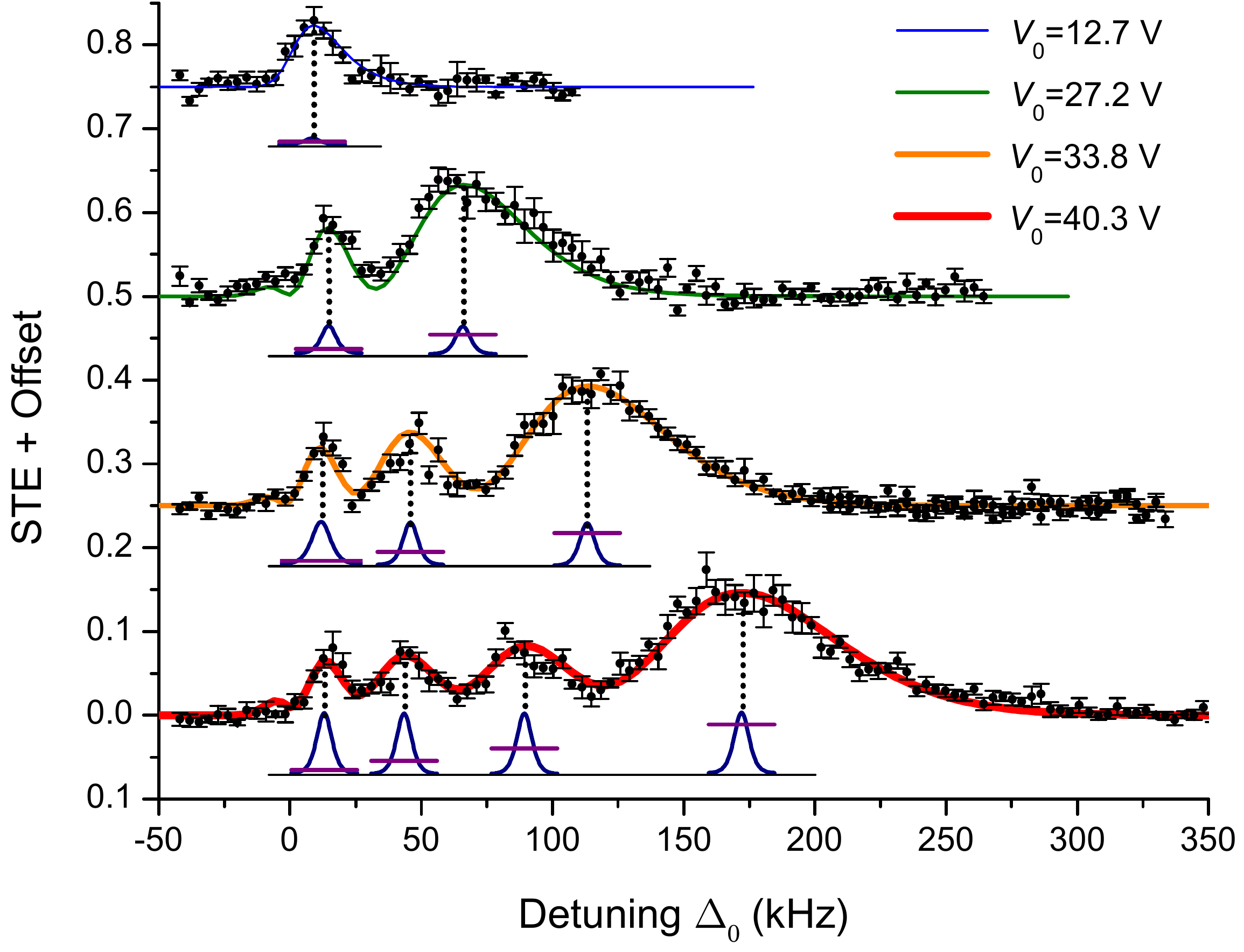}
\caption{
%(color online) Level crossing spectrum with polarizability effect evident. Data shown here are for crossing E, where the small value of $d$ necessitates the use of large $\mathcal{E}$-fields. The curves, shown with optimized values of $d$ and $\alpha$, include effects from finite spatial and velocity distributions, and from imperfect IR centration. Insets show the evolution of the crossing level energies as a function of time, when the $\mathcal{E} = 0$ detuning $\Delta_0$ is set at a peak in the STE data. The approximations yielding Airy function solutions are most valid towards the bottom and right of the plot.
(color online) Level crossing spectra with polarizability effect evident. Data are shown for crossing E,  where the small value of $d$ leads to a pulse area $\Theta  \sim  2d\mathcal{E}_0\sigma_t  \ll  1$ even at large $\mathcal{E}$.  Curves, calculated with optimized values of $d$ and $\alpha$, include effects from finite spatial and velocity distributions and imperfect IR centration. Insets show the evolution of the crossing level energies vs.\ time during the pulse.
}
\label{Fig4}
\end{figure}
Consider a pulse $\Esca(t) = \Esca_0 \exp{[-(t/\sigma_t)^2]}$ and a total detuning $\Delta(t) \cong \Delta^\prime + \beta t^2$, where $\Delta^\prime = \Delta_0(\Bsca) - \Delta_{St}(\Esca)$ and $\beta = 2\Delta_{St}/\sigma_t^2$; here $\Delta_{St} \equiv -\alpha\Esca_0^2/2$ is the maximum Stark shift.  When $|\Delta^\prime | \ll |\Delta_{St}|$, the state amplitude after the pulse is
\begin{equation}
%c_+ = 2\pi \frac{\Dsca \Esca_0}{\beta^{1/3}}~ \mathrm{exp}\left( {\frac{\Delta^\prime
% {\sigma_t^2\beta}    +  \frac{2}{3\sigma_t^6 \beta^2} \right) \mathrm{Ai}\left( 
% \frac{\Delta^\prime}{\beta^{1/3}} +  \frac{1}{\sigma_t^4 \beta^{4/3}} \right).
c_+ \propto \mathrm{exp}\left( \frac{\Phi_{\Delta_0}}{\Phi_S} \right) \mathrm{Ai}\left[ \Phi_S^{-1/3} \left( \Phi_{\Delta_0} -\frac{\Phi_S}{2} +\frac{1}{\Phi_S}\right) \right].
%c_+ \propto \mathrm{exp}\left( \frac{\Delta^\prime}{\sigma_t^2\beta} \right) \mathrm{Ai}\left( \frac{\Delta^\prime}{\beta^{1/3}} +  \frac{1}{\sigma_t^4 \beta^{4/3}} \right),
\end{equation}
Here $\Phi_{\Delta_0} = \Delta_0 \sigma_t$ and $\Phi_S =  2\Delta_{St} \sigma_t$ are measures of the phase accumulated by the states during the pulse due to the unperturbed energy difference and the Stark shift, respectively.  The oscillations in the spectrum described by the Airy function arise because this system is a type of Landau-Zener-St\"{u}ckelberg (LZS) interferometer \cite{Shevchenko10,Mark07}.  The time-varying Stark shift can cause the levels to cross at two values of $t$ (for $\alpha  <  0$  as here, this requires $0  <  \Delta_0  <  -\Delta_{St}$). At each crossing there is a small LZ transition amplitude, while between crossings the states accumulate differential phase $\Phi$ that depends on $\Delta_0$ and $\Delta_{St}$.  When $\Phi  =  (2n+1)\pi$, the transition amplitudes destructively interfere, yielding a zero in the signal.  Typical lineshapes are shown in Fig.\ 4.  The calculated and measured values of $\alpha$ agree well (Table I).

\begin{table}[b]
\begin{tabular}{c|c|c||c|c||c|c||c|c}
\hline
\multicolumn{3}{ c||}{Crossing}& \multicolumn{2}{ c|| }{$\Bsca_0$-4600}&\multicolumn{1}{ c| }{$|d|$}& \multicolumn{1}{ c|| }{$d$}
& \multicolumn{2}{ c }{$\alpha$} \\
\hline
X & $m$ & $m^\prime$        & Meas.              & Fit                      & Meas.        & Calc.          & Meas.                & Calc. \\
\hline
\hline
A            & 1      & 1                   & 04.841(2)      & 04.777               & 3.36(4)      & -3.42         &                           & -0.09                 \\
B            & 1      & 0                   & 16.136(2)      & 16.050               & 3.40(50)     & -4.70         &                           & -0.21                \\                   
C            & 1      & 0                   &                       &                            &                   & -0.07         &                           &                        \\
D            & 0      & 1                   &                       &                           &                   & -0.02          &                           &                        \\
E            & 0      & 0                   & 21.259(2)       & 21.240              & 0.114(5)    & -0.15          & -0.20(2)            & -0.21                 \\                  
F            & 0      & 0                   & 28.214(2)       & 28.278              & 3.53(3)      & -3.50          &                           & -0.09                   \\                    
G           & 0      & -1                  & 38.671(2)       & 38.667               & 0.60(10)    & -0.96          &                           & -0.10                    \\                   
H           & 0      & -1                  & 40.069(2)       & 40.178                & 3.40(40)    & -4.50          & -0.21(2)             & -0.21                 \\
\end{tabular}
\caption{Comparison of data with calculations for all level crossings in $^{138}$BaF. $m (m^\prime)$ is associated with  $|\psi^+_{\uparrow}\rangle  (|\psi^-_{\downarrow}\rangle )$.  Values for $\Bsca_0$ are in Gauss; here ``Fit'' indicates that $g$-factors were optimized, with best values $g_{rot}  =  -0.048$, $g_\parallel  =  -2.00197$, and $(g_\parallel - g_\perp)  =  -0.00594$. Values for $d$ are in kHz/(V/cm); for $\alpha$ in kHz/(V/cm)$^2$. Missing entries were not determined. \label{Table01}}
\end{table}

After accounting for shifts due to $\alpha$, we determine values of $\Bsca_0$ for all observed level crossings and compare them to values calculated by numerical diagonalization of $H_0+H_Z$.  The parameters $g_{rot}$ and $g_{\perp}$ were varied to optimize the agreement (Table I).  The r.m.s.\ deviation of measured and calculated values is $\sigma_{\Bsca_0}/\Bsca_0 \approx  15$ ppm.  This is substantially outside the range of experimental error and of uncertainties from the molecular constants (each $\approx  0.5$ ppm).  Allowing the values of $b, c$ to vary far outside their stated range of uncertainty \cite{Ernst86} dramatically improved the fit (to $\sigma_\Bsca/\Bsca  \approx  0.5$ ppm). We suspect this may indicate the need for additional terms in $H_Z$, as discussed for the case of $^2\Pi$ states in Ref.\ \cite{Tamassia02}; however, such an analysis is outside the scope of this paper.

Finally, we consider the implications of these results for  measurements of PV interactions.  The energy resolution achieved here makes it possible to estimate the sensitivity of such an experiment. The matrix element $W$ of a NSD-PV interaction that connects crossing levels can be measured with statistical uncertainty as small as $\delta W = 1/(2T\sqrt{N_{tot}})$, where $T \approx 150~\mu$s is the interaction time and $N_{tot}$ is the total number of detectable molecules \cite{DeMille08b}.  Our observed count rate $dN/dt = \mathcal{R} N_0 \approx 500$/s corresponds to $\delta W \approx 2\pi\times 0.1$ Hz in 24 hours of integration.  This would be sufficient to measure, with  $\leq 20\%$ precision, the NSD-PV effect predicted in at least 20 molecular species for their relevant isotopes; e.g. in $^{137}$BaF, $W \approx 2\pi\times 5$ Hz \cite{DeMille08b,Borschevsky13}.  
We also note the implications of our slightly imperfect understanding of the full Hamiltonian for this system.  From a perturbative description of the molecular states, we find that any small additional terms in $H_Z$ that might be needed to explain the deviations between observed and predicted level crossing positions could affect the molecular eigenstate composition
% admixtures of basis states with amplitude $\le \sqrt{\delta/\Delta}$, where here $\delta$ is the deviation of the state position from its calculated value and $\Delta$ is the splitting between the state and its nearest neighbor
by no more than 10\% (and typically much less).  Hence such additional terms, although sufficient to shift the positions of level crossings well beyond our resolution, cannot cause changes in the NSD-PV matrix element $W$ larger than this. This level of uncertainty is comparable to that expected in any case from imperfect knowledge of electronic and nuclear structure \cite{DeMille08b}.  Nevertheless, an understanding of the deviations between observed and predicted level crossing positions would of course be desirable, and will be the subject of future investigations.  Altogether, the results reported here demonstrate the level of understanding and control needed to exploit this promising type of system for the study of NSD-PV effects such as those due to nuclear anapole moments \cite{DeMille08b}.

%Acknowledgements
This work was supported by NSF Grant 0457039. 

\bibliography{BaF_level_crossing_2013_v8_noURLs}

\begin{thebibliography}{29}
\expandafter\ifx\csname natexlab\endcsname\relax\def\natexlab#1{#1}\fi
\expandafter\ifx\csname bibnamefont\endcsname\relax
  \def\bibnamefont#1{#1}\fi
\expandafter\ifx\csname bibfnamefont\endcsname\relax
  \def\bibfnamefont#1{#1}\fi
\expandafter\ifx\csname citenamefont\endcsname\relax
  \def\citenamefont#1{#1}\fi
\expandafter\ifx\csname url\endcsname\relax
  \def\url#1{\texttt{#1}}\fi
\expandafter\ifx\csname urlprefix\endcsname\relax\def\urlprefix{URL }\fi
\providecommand{\bibinfo}[2]{#2}
\providecommand{\eprint}[2][]{\url{#2}}

\bibitem[{\citenamefont{Flambaum and Khriplovich}(1985)}]{Flambaum85}
\bibinfo{author}{\bibfnamefont{V.}~\bibnamefont{Flambaum}} \bibnamefont{and}
  \bibinfo{author}{\bibfnamefont{I.}~\bibnamefont{Khriplovich}},
  \bibinfo{journal}{Phys. Lett. A} \textbf{\bibinfo{volume}{110}},
  \bibinfo{pages}{121} (\bibinfo{year}{1985}).

\bibitem[{\citenamefont{Kozlov et~al.}(1991)\citenamefont{Kozlov, Labzovskii,
  and Mitrushchenkov}}]{Kozlov91}
\bibinfo{author}{\bibfnamefont{M.}~\bibnamefont{Kozlov}},
  \bibinfo{author}{\bibfnamefont{L.}~\bibnamefont{Labzovskii}},
  \bibnamefont{and}
  \bibinfo{author}{\bibfnamefont{A.}~\bibnamefont{Mitrushchenkov}},
  \bibinfo{journal}{Sov. Phys. JETP} \textbf{\bibinfo{volume}{73}},
  \bibinfo{pages}{415} (\bibinfo{year}{1991}).

\bibitem[{\citenamefont{DeMille et~al.}(2008)\citenamefont{DeMille, Cahn,
  Murphree, Rahmlow, and Kozlov}}]{DeMille08b}
\bibinfo{author}{\bibfnamefont{D.}~\bibnamefont{DeMille}},
  \bibinfo{author}{\bibfnamefont{S.~B.} \bibnamefont{Cahn}},
  \bibinfo{author}{\bibfnamefont{D.}~\bibnamefont{Murphree}},
  \bibinfo{author}{\bibfnamefont{D.~A.} \bibnamefont{Rahmlow}},
  \bibnamefont{and} \bibinfo{author}{\bibfnamefont{M.~G.}
  \bibnamefont{Kozlov}}, \bibinfo{journal}{Phys. Rev. Lett.}
  \textbf{\bibinfo{volume}{100}}, \bibinfo{pages}{023003}
  (\bibinfo{year}{2008}).

\bibitem[{\citenamefont{Nguyen et~al.}(1997)\citenamefont{Nguyen, Budker,
  DeMille, and Zolotorev}}]{Nguyen97}
\bibinfo{author}{\bibfnamefont{A.~T.} \bibnamefont{Nguyen}},
  \bibinfo{author}{\bibfnamefont{D.}~\bibnamefont{Budker}},
  \bibinfo{author}{\bibfnamefont{D.}~\bibnamefont{DeMille}}, \bibnamefont{and}
  \bibinfo{author}{\bibfnamefont{M.}~\bibnamefont{Zolotorev}},
  \bibinfo{journal}{Phys. Rev. A} \textbf{\bibinfo{volume}{56}},
  \bibinfo{pages}{3453} (\bibinfo{year}{1997}).

\bibitem[{\citenamefont{Haxton and Wieman}(2001)}]{Haxton01}
\bibinfo{author}{\bibfnamefont{W.~C.} \bibnamefont{Haxton}} \bibnamefont{and}
  \bibinfo{author}{\bibfnamefont{C.~E.} \bibnamefont{Wieman}},
  \bibinfo{journal}{Annu. Rev. Nucl. Part. Sci.} \textbf{\bibinfo{volume}{51}},
  \bibinfo{pages}{261} (\bibinfo{year}{2001}).

\bibitem[{\citenamefont{Flambaum and Murray}(1997)}]{Flambaum97}
\bibinfo{author}{\bibfnamefont{V.~V.} \bibnamefont{Flambaum}} \bibnamefont{and}
  \bibinfo{author}{\bibfnamefont{D.~W.} \bibnamefont{Murray}},
  \bibinfo{journal}{Phys. Rev. C} \textbf{\bibinfo{volume}{56}},
  \bibinfo{pages}{1641} (\bibinfo{year}{1997}).

\bibitem[{\citenamefont{Wallis et~al.}(2009)\citenamefont{Wallis, Gardiner, and
  Hutson}}]{Wallis09}
\bibinfo{author}{\bibfnamefont{A.~O.~G.} \bibnamefont{Wallis}},
  \bibinfo{author}{\bibfnamefont{S.~A.} \bibnamefont{Gardiner}},
  \bibnamefont{and} \bibinfo{author}{\bibfnamefont{J.~M.}
  \bibnamefont{Hutson}}, \bibinfo{journal}{Phys. Rev. Lett.}
  \textbf{\bibinfo{volume}{103}}, \bibinfo{pages}{083201}
  (\bibinfo{year}{2009}).

\bibitem[{\citenamefont{P\'{e}rez-Rios
  et~al.}(2010)\citenamefont{P\'{e}rez-Rios, Herrera, and
  Krems}}]{Perez-Rios10}
\bibinfo{author}{\bibfnamefont{J.}~\bibnamefont{P\'{e}rez-Rios}},
  \bibinfo{author}{\bibfnamefont{F.}~\bibnamefont{Herrera}}, \bibnamefont{and}
  \bibinfo{author}{\bibfnamefont{R.~V.} \bibnamefont{Krems}},
  \bibinfo{journal}{New Journal of Physics} \textbf{\bibinfo{volume}{12}},
  \bibinfo{pages}{103007} (\bibinfo{year}{2010}).

\bibitem[{\citenamefont{Alyabyshev et~al.}(2012)\citenamefont{Alyabyshev,
  Lemeshko, and Krems}}]{Alyabyshev12}
\bibinfo{author}{\bibfnamefont{S.~V.} \bibnamefont{Alyabyshev}},
  \bibinfo{author}{\bibfnamefont{M.}~\bibnamefont{Lemeshko}}, \bibnamefont{and}
  \bibinfo{author}{\bibfnamefont{R.~V.} \bibnamefont{Krems}},
  \bibinfo{journal}{Phys. Rev. A} \textbf{\bibinfo{volume}{86}},
  \bibinfo{pages}{013409} (\bibinfo{year}{2012}).

\bibitem[{\citenamefont{Herzberg}(1950)}]{Herzberg50}
\bibinfo{author}{\bibfnamefont{G.}~\bibnamefont{Herzberg}},
  \emph{\bibinfo{title}{Molecular Spectra and Molecular Structure: I. Spectra
  of Diatomic Molecules, 2nd Ed.}} (\bibinfo{publisher}{D. Van Nostrand
  Company, Inc}, \bibinfo{address}{New York}, \bibinfo{year}{1950}).

\bibitem[{\citenamefont{Childs}(1992)}]{Childs92}
\bibinfo{author}{\bibfnamefont{W.}~\bibnamefont{Childs}},
  \bibinfo{journal}{Physics Reports} \textbf{\bibinfo{volume}{211}},
  \bibinfo{pages}{113 } (\bibinfo{year}{1992}), ISSN \bibinfo{issn}{0370-1573}.

\bibitem[{\citenamefont{Brown and Carrington}(2003)}]{Brown03}
\bibinfo{author}{\bibfnamefont{J.~M.} \bibnamefont{Brown}} \bibnamefont{and}
  \bibinfo{author}{\bibfnamefont{A.}~\bibnamefont{Carrington}},
  \emph{\bibinfo{title}{Rotational Spectroscopy of Diatomic Molecules}}
  (\bibinfo{publisher}{Cambridge University Press},
  \bibinfo{address}{Cambridge}, \bibinfo{year}{2003}).

\bibitem[{\citenamefont{Ryzlewicz and T\"{o}rring}(1980)}]{Ryzlewicz80}
\bibinfo{author}{\bibfnamefont{C.}~\bibnamefont{Ryzlewicz}} \bibnamefont{and}
  \bibinfo{author}{\bibfnamefont{T.}~\bibnamefont{T\"{o}rring}},
  \bibinfo{journal}{Chem. Phys.} \textbf{\bibinfo{volume}{51}},
  \bibinfo{pages}{329 } (\bibinfo{year}{1980}), ISSN \bibinfo{issn}{0301-0104}.

\bibitem[{\citenamefont{Ryzlewicz et~al.}(1982)\citenamefont{Ryzlewicz,
  Sch\"{u}tze-Pahlmann, Hoeft, and T\"{o}rring}}]{Ryzlewicz82}
\bibinfo{author}{\bibfnamefont{C.}~\bibnamefont{Ryzlewicz}},
  \bibinfo{author}{\bibfnamefont{H.-U.} \bibnamefont{Sch\"{u}tze-Pahlmann}},
  \bibinfo{author}{\bibfnamefont{J.}~\bibnamefont{Hoeft}}, \bibnamefont{and}
  \bibinfo{author}{\bibfnamefont{T.}~\bibnamefont{T\"{o}rring}},
  \bibinfo{journal}{Chem. Phys.} \textbf{\bibinfo{volume}{71}},
  \bibinfo{pages}{389 } (\bibinfo{year}{1982}), ISSN \bibinfo{issn}{0301-0104}.

\bibitem[{\citenamefont{Ernst et~al.}(1986)\citenamefont{Ernst, K\"{a}ndler,
  and T\"{o}rring}}]{Ernst86}
\bibinfo{author}{\bibfnamefont{W.~E.} \bibnamefont{Ernst}},
  \bibinfo{author}{\bibfnamefont{J.}~\bibnamefont{K\"{a}ndler}},
  \bibnamefont{and}
  \bibinfo{author}{\bibfnamefont{T.}~\bibnamefont{T\"{o}rring}},
  \bibinfo{journal}{J. Chem. Phys.} \textbf{\bibinfo{volume}{84}},
  \bibinfo{pages}{4769} (\bibinfo{year}{1986}).

\bibitem[{\citenamefont{Weltner}(1983)}]{Weltner83}
\bibinfo{author}{\bibfnamefont{W.}~\bibnamefont{Weltner}},
  \emph{\bibinfo{title}{Magnetic Atoms and Molecules}} (\bibinfo{publisher}{Van
  Nostrand Reinhold}, \bibinfo{address}{New York}, \bibinfo{year}{1983}).

\bibitem[{\citenamefont{Curl~Jr.}(1965)}]{Curl65}
\bibinfo{author}{\bibfnamefont{R.}~\bibnamefont{Curl~Jr.}},
  \bibinfo{journal}{Mol. Phys.} \textbf{\bibinfo{volume}{9}},
  \bibinfo{pages}{585} (\bibinfo{year}{1965}).

\bibitem[{\citenamefont{Stone}(2005)}]{Stone05}
\bibinfo{author}{\bibfnamefont{N.}~\bibnamefont{Stone}},
  \bibinfo{journal}{Atomic Data and Nuclear Data Tables}
  \textbf{\bibinfo{volume}{90}}, \bibinfo{pages}{75 } (\bibinfo{year}{2005}),
  ISSN \bibinfo{issn}{0092-640X}.

\bibitem[{\citenamefont{Wieman et~al.}(1987)\citenamefont{Wieman, Noecker,
  Masterson, and Cooper}}]{Wieman87}
\bibinfo{author}{\bibfnamefont{C.~E.} \bibnamefont{Wieman}},
  \bibinfo{author}{\bibfnamefont{M.~C.} \bibnamefont{Noecker}},
  \bibinfo{author}{\bibfnamefont{B.~P.} \bibnamefont{Masterson}},
  \bibnamefont{and} \bibinfo{author}{\bibfnamefont{J.}~\bibnamefont{Cooper}},
  \bibinfo{journal}{Phys. Rev. Lett.} \textbf{\bibinfo{volume}{58}},
  \bibinfo{pages}{1738} (\bibinfo{year}{1987}).

\bibitem[{\citenamefont{Stalnaker et~al.}(2006)\citenamefont{Stalnaker, Budker,
  Freedman, Guzman, Rochester, and Yashchuk}}]{Stalnaker06}
\bibinfo{author}{\bibfnamefont{J.~E.} \bibnamefont{Stalnaker}},
  \bibinfo{author}{\bibfnamefont{D.}~\bibnamefont{Budker}},
  \bibinfo{author}{\bibfnamefont{S.~J.} \bibnamefont{Freedman}},
  \bibinfo{author}{\bibfnamefont{J.~S.} \bibnamefont{Guzman}},
  \bibinfo{author}{\bibfnamefont{S.~M.} \bibnamefont{Rochester}},
  \bibnamefont{and} \bibinfo{author}{\bibfnamefont{V.~V.}
  \bibnamefont{Yashchuk}}, \bibinfo{journal}{Phys. Rev. A}
  \textbf{\bibinfo{volume}{73}}, \bibinfo{pages}{043416}
  (\bibinfo{year}{2006}).

\bibitem[{\citenamefont{Hioe and Carroll}(1985)}]{Hioe85}
\bibinfo{author}{\bibfnamefont{F.~T.} \bibnamefont{Hioe}} \bibnamefont{and}
  \bibinfo{author}{\bibfnamefont{C.~E.} \bibnamefont{Carroll}},
  \bibinfo{journal}{Phys. Rev. A} \textbf{\bibinfo{volume}{32}},
  \bibinfo{pages}{1541} (\bibinfo{year}{1985}).

\bibitem[{\citenamefont{Allen and Eberly}(1987)}]{Allen87}
\bibinfo{author}{\bibfnamefont{L.}~\bibnamefont{Allen}} \bibnamefont{and}
  \bibinfo{author}{\bibfnamefont{J.~H.} \bibnamefont{Eberly}},
  \emph{\bibinfo{title}{Optical Resonance and Two-Level Atoms}}
  (\bibinfo{publisher}{Dover Publications}, \bibinfo{address}{New York},
  \bibinfo{year}{1987}).

\bibitem[{\citenamefont{Effantin et~al.}(1990)\citenamefont{Effantin, Bernard,
  d'Incan, Wannous, Verg\'{e}s, and Barrow}}]{Effantin90}
\bibinfo{author}{\bibfnamefont{C.}~\bibnamefont{Effantin}},
  \bibinfo{author}{\bibfnamefont{A.}~\bibnamefont{Bernard}},
  \bibinfo{author}{\bibfnamefont{J.}~\bibnamefont{d'Incan}},
  \bibinfo{author}{\bibfnamefont{G.}~\bibnamefont{Wannous}},
  \bibinfo{author}{\bibfnamefont{J.}~\bibnamefont{Verg\'{e}s}},
  \bibnamefont{and} \bibinfo{author}{\bibfnamefont{R.}~\bibnamefont{Barrow}},
  \bibinfo{journal}{Mol. Phys.} \textbf{\bibinfo{volume}{70}},
  \bibinfo{pages}{735} (\bibinfo{year}{1990}).

\bibitem[{\citenamefont{Murphree et~al.}(2007)\citenamefont{Murphree, Cahn,
  Rahmlow, and DeMille}}]{Murphree07}
\bibinfo{author}{\bibfnamefont{D.}~\bibnamefont{Murphree}},
  \bibinfo{author}{\bibfnamefont{S.~B.} \bibnamefont{Cahn}},
  \bibinfo{author}{\bibfnamefont{D.}~\bibnamefont{Rahmlow}}, \bibnamefont{and}
  \bibinfo{author}{\bibfnamefont{D.}~\bibnamefont{DeMille}},
  \bibinfo{journal}{J. Magn. Reson.} \textbf{\bibinfo{volume}{188}},
  \bibinfo{pages}{160} (\bibinfo{year}{2007}).

\bibitem[{\citenamefont{Rosen and Zener}(1932)}]{Rosen32}
\bibinfo{author}{\bibfnamefont{N.}~\bibnamefont{Rosen}} \bibnamefont{and}
  \bibinfo{author}{\bibfnamefont{C.}~\bibnamefont{Zener}},
  \bibinfo{journal}{Phys. Rev.} \textbf{\bibinfo{volume}{40}},
  \bibinfo{pages}{502} (\bibinfo{year}{1932}).

\bibitem[{\citenamefont{Shevchenko et~al.}(2010)\citenamefont{Shevchenko,
  Ashhab, and Nori}}]{Shevchenko10}
\bibinfo{author}{\bibfnamefont{S.}~\bibnamefont{Shevchenko}},
  \bibinfo{author}{\bibfnamefont{S.}~\bibnamefont{Ashhab}}, \bibnamefont{and}
  \bibinfo{author}{\bibfnamefont{F.}~\bibnamefont{Nori}},
  \bibinfo{journal}{Physics Reports} \textbf{\bibinfo{volume}{492}},
  \bibinfo{pages}{1 } (\bibinfo{year}{2010}), ISSN \bibinfo{issn}{0370-1573}.

\bibitem[{\citenamefont{Mark et~al.}(2007)\citenamefont{Mark, Kraemer,
  Waldburger, Herbig, Chin, N\"{a}gerl, and Grimm}}]{Mark07}
\bibinfo{author}{\bibfnamefont{M.}~\bibnamefont{Mark}},
  \bibinfo{author}{\bibfnamefont{T.}~\bibnamefont{Kraemer}},
  \bibinfo{author}{\bibfnamefont{P.}~\bibnamefont{Waldburger}},
  \bibinfo{author}{\bibfnamefont{J.}~\bibnamefont{Herbig}},
  \bibinfo{author}{\bibfnamefont{C.}~\bibnamefont{Chin}},
  \bibinfo{author}{\bibfnamefont{H.-C.} \bibnamefont{N\"{a}gerl}},
  \bibnamefont{and} \bibinfo{author}{\bibfnamefont{R.}~\bibnamefont{Grimm}},
  \bibinfo{journal}{Phys. Rev. Lett.} \textbf{\bibinfo{volume}{99}},
  \bibinfo{pages}{113201} (\bibinfo{year}{2007}).

\bibitem[{\citenamefont{Tamassia et~al.}(2002)\citenamefont{Tamassia, Brown,
  and Watson}}]{Tamassia02}
\bibinfo{author}{\bibfnamefont{F.}~\bibnamefont{Tamassia}},
  \bibinfo{author}{\bibfnamefont{J.~M.} \bibnamefont{Brown}}, \bibnamefont{and}
  \bibinfo{author}{\bibfnamefont{J.~K.} \bibnamefont{Watson}},
  \bibinfo{journal}{Mol. Phys.} \textbf{\bibinfo{volume}{100}},
  \bibinfo{pages}{3485} (\bibinfo{year}{2002}).

\bibitem[{\citenamefont{Borschevsky et~al.}(2013)\citenamefont{Borschevsky,
  Ilia$\check{s}$, Dzuba, Flambaum, and Schwerdtfeger}}]{Borschevsky13}
\bibinfo{author}{\bibfnamefont{A.}~\bibnamefont{Borschevsky}},
  \bibinfo{author}{\bibfnamefont{M.}~\bibnamefont{Ilia$\check{s}$}},
  \bibinfo{author}{\bibfnamefont{V.~A.} \bibnamefont{Dzuba}},
  \bibinfo{author}{\bibfnamefont{V.~V.} \bibnamefont{Flambaum}},
  \bibnamefont{and}
  \bibinfo{author}{\bibfnamefont{P.}~\bibnamefont{Schwerdtfeger}},
  \bibinfo{journal}{Phys. Rev. A} \textbf{\bibinfo{volume}{88}},
  \bibinfo{pages}{022125} (\bibinfo{year}{2013}).

\end{thebibliography}

%%%%%%%%%%%%%%%%%%%%%%%%%%%%%%%%%%%%%%%%%%%%%%%%%%%%%%%%%%%%%%%%%%%%%%%%%%%%%%%%%%%%%%%%%%%%%%%
\end{document}